\documentclass[11pt]{article}
\usepackage{graphicx}
\usepackage{moriond,epsfig}

\def\Journal#1#2#3#4{{#1} {\bf #2}, #3 (#4)}

\def\PRL{\em Phys. Rev. Lett.}
\def\PRD{{\em Phys. Rev.} D}

\def\CQG{\em Class. Quant. Grav.}
\def\APJ{\em The Astrophysical Journal}

\def\be{\begin{equation}}
\def\ee{\end{equation}}
\def\bea{\begin{eqnarray}}
\def\eea{\end{eqnarray}}
\begin{document}
\vspace*{4cm}
\title{SEARCHES FOR CONTINUOUS GRAVITATIONAL WAVE SIGNALS AND STOCHASTIC BACKGROUNDS IN LIGO AND VIRGO DATA}

\author{C. PALOMBA for the LIGO Scientific Collaboration and the Virgo Collaboration}

\address{Istituto Nazionale di Fisica Nucleare, sezione di Roma, I-00185 Roma, Italy}


\maketitle\abstract{
We present results from searches of recent LIGO and Virgo data for continuous gravitational wave signals (CW) from spinning neutron stars and for a stochastic gravitational wave background (SGWB). 

The first part of the talk is devoted to CW analysis with a focus on two types of searches. In the targeted search of known neutron stars a precise knowledge of the star parameters is used to apply optimal filtering methods. In the absence of a signal detection, in a few cases, an upper limit on strain amplitude can be set that beats the spindown limit derived from attributing spin-down energy loss to the emission of gravitational waves. 
In contrast, blind all-sky searches are not directed at specific sources, but rather explore as large a portion of the parameter space as possible. 
Fully coherent methods cannot be used for these kind of searches which pose a non trivial computational challenge. 

The second part of the talk is focused on SGWB searches. A stochastic background of gravitational waves is expected to be produced by the superposition of many incoherent sources of cosmological or astrophysical origin. Given the random nature of this kind of signal, it is not possible to distinguish it from noise using a single detector. A typical data analysis strategy relies on cross-correlating the data from a pair or several pairs of detectors, which allows discriminating the searched signal from instrumental noise. 

Expected sensitivities and prospects for detection from the next generation of interferometers are also discussed for both kind of sources. 
}

\section{Introduction}
The most recent results obtained in the search of CW and SGWB have used data from LIGO S5~\cite{s5} and Virgo VSR2~\cite{v2} runs.
S5 run involved all three LIGO detectors, Hanford 4km (H1), Hanford 2km (H2) and Livingston 4km (L1), and took place from November 2005 to September 2007 with an average single-interferometer duty cycle of 73.6$\%$, an average two-site coincident duty cycle of 59.4$\%$ and an average triple-interferometer duty cycle of 52.5$\%$. Virgo (V1) VSR2 run took place from July 2009 to January 2010 with a duty cycle of 80.4$\%$. 
At low frequency, say below 70~Hz, VSR2 sensitivity was better than S5. At intermediate frequency, between 70~Hz and 500~Hz, S5 sensitivity was better than VSR2. At frequency above about 500~Hz the sensitivity of the two runs were very similar.

In 2010 two more scientific runs, LIGO S6 and Virgo VSR3, took place. The data are being analyzed and some interesting results have already been obtained, but are still under internal review, so we will not discuss them here.

A generic gravitational wave (GW) signal is described by a tensor metric perturbation ${\bf h}(t)=h_+(t)\, {\bf e_+} + h_{\times}(t)\, {\bf e_{\times}},$ where ${\bf e_+}$ and ${\bf e_{\times}}$ are the two basis polarization tensors. The form of the two amplitudes depends on the specific kind of signal.

\section{The search for continuous gravitational wave signals}
Rapidly spinning neutron stars, isolated or in binary systems, are a potential source of CW. To emit GW some degree of non-axisymmetry is required. It can be due to several mechanisms including elastic stress or magnetic field which induce a deformation of the neutron star shape, free-precession around the rotation axis or accretion of matter from a companion star. 
The size of the distortion, typically measured by the {\it ellipticity} $\epsilon=\frac{I_{xx}-I_{yy}}{I_{zz}}$, which is defined in terms of the star principal moments of inertia, can provide important information on the neutron star equation of state.

The signal emitted by a tri-axial neutron star rotating around a principal axis of inertia is characterized by amplitudes
\be
h_+(t) = h_0\left(\frac{1+\cos^2{\iota}}{2}\right)\cos{\Phi(t)};~~~~
h_{\times}(t) = h_0\cos{\iota}\sin{\Phi(t)},
\label{eq:hpluscross}
\ee
The angle $\iota$ is the inclination of the star's rotation axis with respect to the line of sight and $\Phi(t)$ is the signal phase function, where $t$ is the detector time, while
the amplitude $h_0$ is given by
\begin{equation}
h_0=\frac{4\pi^2G}{c^4}\frac{I_{zz}\epsilon f^2}{d},
\label{eq:h0}
\end{equation} 
being $d$ the star distance and $f$ the signal frequency (twice the star rotation frequency). While we expect $f<2$~kHz and $d<$10~kpc, the typical value of the ellipticity is largely unknown. Standard equations of state (EOS) of neutron star matter foresee maximum value of the ellipticity~\cite{horo} of the order of $\epsilon_{max}\approx 5\cdot 10^{-6}$. For some 'exotic' EOS a maximum value $\epsilon_{max}\approx 10^{-2}-10^{-4}$ is foreseen~\cite{owen}$^,$~\cite{lin}$^,$~\cite{hask}. 

The signal frequency gradually decreases due to the intrinsic source spin-down, caused by elecromagnetic and hopefully gravitational energy losses. The received signal phase is affected by the Doppler modulation due to the detector-source relative motion and by some relativistic effects. Moreover, the signal is also affected by the amplitude and phase modulation due to the detector beam-pattern functions $F_+(t;\psi),~F_{\times}(t;\psi)$,
which depend on the polarization angle $\psi$, on the source position in the sky and on the detector position and orientation on the Earth. 

Assuming that the observed spin-down $\dot{f}$ of a given neutron star is totally due to the emission of GW, an absolute upper limit to the amplitude of the GW signal, called {\it spin-down limit}, can be derived~\cite{ab1}:
\begin{equation}
h^{sd}_0=8.06\cdot 10^{-19}\,I_{38}\,d^{-1}_{\rm kpc}\,\sqrt{\frac{|(\dot{f}/{\rm Hz}\,{\rm s}^{-1})|}{(f/{\rm Hz})}},
\label{eq:hsd}
\end{equation}
where $I_{38}$ is the star's moment of inertia in units of
$10^{38}$\,kg\,m$^2$ and $d_{\rm kpc}$ is the star's distance from the Sun in kiloparsecs.
Going below the {\it spin-down limit} means we are putting a constraint on the fraction of spin-down energy due to the emission of GW.\newline

Two types of CW searches have received the most effort up to now: {\it targeted} searches and {\it wide parameter} searches.

In the targeted searches the source parameters $(\alpha,\delta,f,\dot{f},...)$ are assumed to be known with high accuracy. The search for known pulsars belongs to this category. This kind of search is computationally cheap and a fully coherent analysis, based on matched filtering, over long observation time is feasible. 
Various methods of implementing matched filtering have been developed~\cite{dupu}$^,$~\cite{krola}$^,$~\cite{asto}. In order to make a coherent analysis over long times Doppler, Einstein and possibly Shapiro effects must be accurately compensated.
Radio-astronomic observations can be used to accurately track the GW signal phase evolution (assuming the GW signal is phase locked to the EM pulses). Moreover, they are also important to know if a {\it glitch}, i.e. a sudden jump in frequency and frequency derivative, occurred during the period of data to be analyzed.  

The sensitivity  of a coherent search, i.e. the minimum signal amplitude detectable over an observation time $T_{obs}$, with a false alarm probability of 1~$\%$ and a false dismissal probability of 10~$\%$ and taking also an average over source and detector parameters, is given by
\begin{equation}
h_{0,min}\approx 11\sqrt{\frac{S_n(f)}{T_{obs}}}
\label{snr_cohe}
\end{equation}
where $S_n(f)$ is the detector noise spectrum. The exact value of the coefficient depends on the specific analysis method employed.

A coherent search for CW using LIGO-S5 data has been recently done for more than 100 pulsars~\cite{ab2} but the resulting upper limits have beaten the spin-down limit for only the Crab pulsar and have grazed it for PSRJ0537-6910 (less than a factor of 2 above). For the Crab the analysis was carried out both assuming the polarization parameters $\iota,\psi$ are unknown (uniform priors) and that they are known with values estimated from x-ray observations~\cite{ng} (restricted priors). Updated ephemeris from Jodrell Bank were used. The $95\%$ degree-of-belief upper limits are $h_0^{95\%}=2.4\cdot 10^{-25}$ (uniform prior) and $h_0^{95\%}=1.9\cdot 10^{-25}$ (restricted prior) corresponding to a star ellipticity of $\sim 10^{-4}$. These results are below the spin-down limit by a factor of about 7, and constrain the fraction of spin-down energy due to the emission of GW to about 2$\%$ (assuming the canonical value for the star moment of inertia, $10^{38}~kg\cdot m^2$).  
We expect to improve the upper limit on the Crab pulsar by jointly analysing data from LIGO S5, S6 runs and Virgo VSR2, VSR3, VSR4 runs (this last tentatively scheduled for summer 2011).

Virgo VSR2 data have been used for a coherent search of CW from the Vela pulsar. Updated ephemeris have been computed using TEMPO2 software from the time-of-arrivals of EM pulses observed by Hobart and Hartebeesthoek radio-telescopes. The excellent seismic isolation of Virgo detector allows for a very good sensitivity at low frequencies thus making the spin-down limit potentially beatable. Results of this analysis are described in~\cite{vela}.\newline

In the wide parameter searches the analysis is done over a portion of the source parameter space as large as possible. In particular, we would like to search for unknown sources located everywhere in the sky, with signal frequency as high as 2~kHz and with values of spin-down as large as possible. This kind of analysis is computationally bound. Fully coherent methods which would allow to reach the `best' search sensitivity, like the ones used for targeted searches, are unfeasible due to the computing power limitation. Various incoherent methods have been developed in which the data are divided in small Fourier transformed segments which are then properly combined to compensate for Doppler and spin-down for a particular source location~\cite{pof}$^-$~\cite{ab4}. In the so-called {\it hierarchical methods} coherent (over relatively short periods) and incoherent steps are alternated in order to increase sensitivity~\cite{bra2}$^,$~\cite{fra}. The output of an analysis is given by a set of {\it candidates}, i.e. points in the source parameter space with high values of a given statistic and which need a deeper study. Typically {\it coincidences} are done among the candidates obtained by the analysis over different data sets in order to reduce the false alarm probability~\cite{c6c7}$^,$~\cite{eh1}. The surviving candidates can be then analyzed coherently over longer times in order to discard them or confirm detection. The basic sensitivity of a wide parameter search is given by
\begin{equation}
h_{0,min}\approx \frac{25}{N^{1/4}}\sqrt{\frac{S_n(f)}{T_{coh}}}
\label{snr_incohe}
\end{equation}  
where $N$ is the number of segments in which the data are divided, each of length $T_{coh}$. The exact value of the numerical factor depends again on the specific incoherent method used and weakly also on the parameter space that is being considered. 

\begin{figure}
\begin{center}
\includegraphics[width=0.8\textwidth]{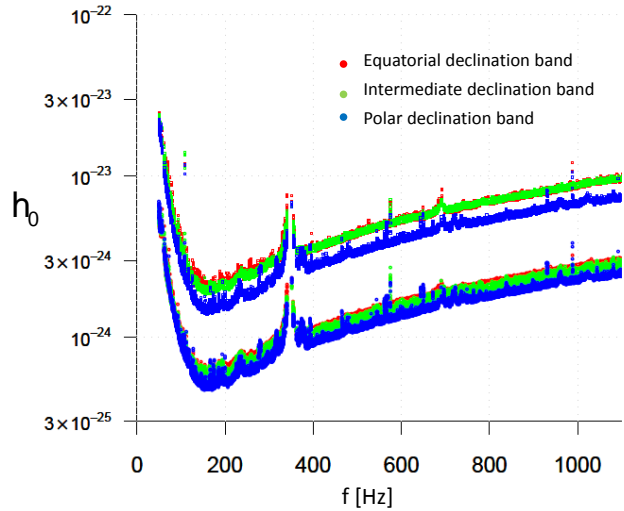}
\caption{Minimum (H1 or L1) $95\%$ confidence level upper limits on signal amplitude for equatorial, intermediate and polar declination bands. Lower curves corresponds to best neutron star orientation, upper curves to worst neutron star orientation.  Figure adapted from PRL~\protect \cite{pfluxS5}.  
\label{fig:S5_powerf}}
\end{center}
\end{figure}

\begin{figure}
\begin{center}
\includegraphics[width=0.8\textwidth]{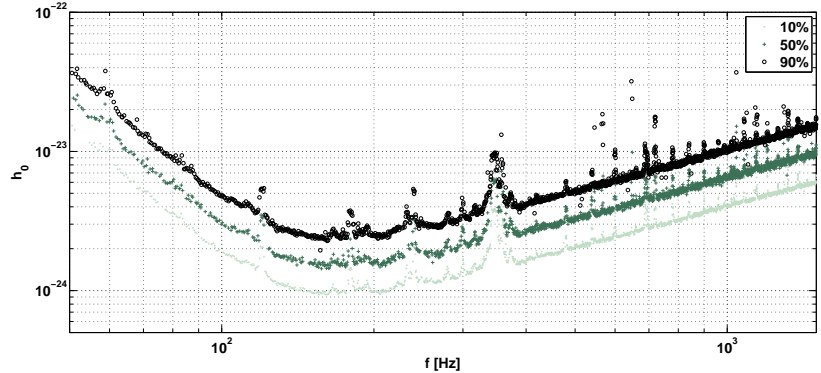}
\caption{Estimated sensitivity of the Einstein@Home search for early LIGO S5 data. The three curves show the source amplitude $h_0$ at which $10\%$ (bottom), $50\%$ (middle), $90\%$ (top) of the simulated sources would be confidently detected. Figure adapted from PRD~\protect \cite{eh1}.  
\label{fig:S5R1}}
\end{center}
\end{figure}

Early LIGO S5-data have been analyzed with two different methods. No GW signal has been detected but interesting upper limits have been placed. A search using the first 8 months of S5 has been described in~\cite{pfluxS5}. It covered the whole sky, the frequency band $50-1100~Hz$ and a range of spin-down values between $-5\cdot 10^{-9}~Hz/s$ and $0$. At the highest frequency the search would have been sensitive to the GW emitted by a neutron star placed at $~500~pc$ with equatorial ellipticity larger than $10^{-6}$. In Fig. \ref{fig:S5_powerf} the upper limits as a function of frequency are shown.
Another search, using the Einstein@Home infrastructure - a volunteer distributed computing project~\cite{ehweb}, was done over the first 2 months of S5~\cite{eh1}. The analysis consisted in matched filtering over 30-hours data segments followed by incoherent combination of results via a concidence strategy. The explored parameter space consisted in the full sky, frequency range $50-1500~Hz$, spin-down range between $-2\cdot 10^{-9}~Hz/s$ and $0$. This search would have been sensitive to $90~\%$ of signals in the frequency band $125-225~Hz$ with amplitude greater than $3\cdot 10^{-24}$. The search sensitivity, estimated through the injection of software simulated signals, is shown in Fig. \ref{fig:S5R1}.

Various improvements to the wide parameter search pipelines are being implemented in order to have a better sensitivity at fixed computing power~\cite{holg}$^,$~\cite{derg2}$^,$~\cite{asto2}.

Other kinds of searches have been or are being developed and applied to detector data: {\it directed} searches, searches for accreting neutron stars, searches for neutron stars in binary systems, transient searches for short-lived signals. In particular, {\it directed} searches are somewhat intermediate between targeted and all-sky. To this category belong, e.g., the search for sources with known position but unknown frequency (like the compact objects in supernova remnants) and the search over relatively small sky area (like the galactic center or globular clusters). An interesting upper limit has been obtained in the analysis of $\sim 12~$ days of S5 data searching for GW signals from the supernova remnant Cassiopeia A~\cite{cassio}. The source position is known and a coherent search over the frequency range $100-300~Hz$ and a wide range of spin-down values   
has been done establishing a $95\%$ confidence upper limit below the indirect limit based on energy conservation and age of the remnant. This search established also the first upper limit on r-modes amplitude.\newline  


With Advanced LIGO and Virgo detectors the spin-down limit on GW emission from known pulsars will be beatable for tens of objects and in few cases the minimum detectable ellipticity will be below $10^{-5}$ and down to $10^{-8}$, a range of values which is sustainable also by standard neutron star EOS. Concerning all-sky searches, nearby {\it gravitars} (say, a few hundreds of parsecs away) would be detectable for ellipticity larger than a few units in $10^{-8}$. Objects with ellipticity of the order of $10^{-6}$ would be detectable up to the Galactic center (see, e.g., Fig. 41 of~\cite{ab4} after up-scaling by a factor of about 10 the distance associated to red curves).  

\section{The search for stochastic gravitational wave backgrounds}
Typically, two kinds of stochastic gravitational wave backgrounds (SGWB) are considered. Cosmological backgrounds, due to processes taking place in the very early stages of Universe evolution, like amplification of vacuum fluctuations, phase transition, cosmic string cusps. These kinds of backgrounds are expected to be stationary, gaussian, unpolarized and, in a first approximation, isotropic. Astrophysical backgrounds, due to the superposition of many unresolved sources, since the beginning of stellar activity, like core collapse to supernovae or the final stages of compact binary mergers. The assumption of isotropy would not hold, if of galactic origin, and an astrophysical background could also be not gaussian, if the number of contributing sources is not very large. Detection of a background of cosmological origin may allow us to probe time scales and energy not accessible with conventional astronomy or accelerators. Even in case of non-detection important constraints to model parameters can be established. An astrophysical background, interesting in its own right, could in fact be a foreground obscuring the cosmological background in some frequency band. See references in the review papers by Maggiore~\cite{magi} and Regimbau~\cite{regi} for a more detailed description of various possible sources of SGWB.\newline

\begin{figure}
\begin{center}
\includegraphics[width=0.6\textwidth]{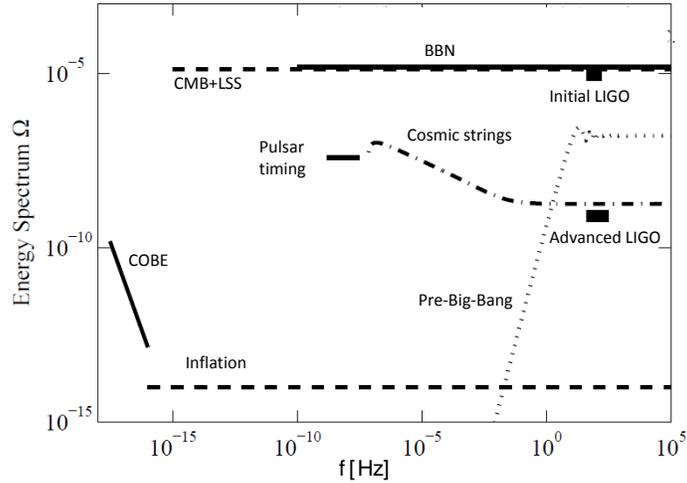}
\caption{Theoretical predictions of various cosmological SGWB and observational bounds. The cosmic super-string curve corresponds to parameters $p=0.1$ (probability that two strings would undergo reconnection to form a loop), $\epsilon=7\cdot 10^{-5}$ (loop size), $G \mu=10^{-8}$ (dimensionless string tension). The Pre-Big-Bang curve corresponds to parameters $\mu=1.5$ (measure of the growth of the dilaton field during the stringy phase), $f_1=4.3\cdot 10^{10}~Hz$ (redshifted frequency of GWs beginning at the end of the stringy phase and lasting to the present day), $f_s=100~Hz$ (redshifted frequency of GWs beginning at the advent of the stringy phase, and lasting to the present day). The bound due to Advanced LIGO is based on its planned sensitivity curve. Figure prepared using the tool at http://homepages.spa.umn.edu/\%7Egwplotter/. 
\label{fig:stoch_new}}
\end{center}
\end{figure}

A SGWB is usually characterized by the dimensionless parameter
\be
\Omega_{GW}(f)=\frac{1}{\rho_c}\frac{d\rho_{gw}}{dln f}
\label{omegw}
\ee
where $\rho_{gw}$ is the gravitational wave densitiy, $f$ is the frequency in the observer frame and $\rho_c=\frac{3H_0^2}{8\pi G}$ is the critical energy density to close the Universe ($H_0$ is the Hubble constant). It is also useful, in particular to make easier the comparison with detector sensitivity, to express the background in terms of the signal energy spectral density
\be
S_h(f)=\frac{3H_0^2}{4\pi^2}\Omega_{GW}(f)
\ee 
In Fig.\ref{fig:stoch_new} theoretical predictions (for given model parameters, see figure caption) of various SGWB of cosmological origin and observational bounds are shown.

The signal would appear as excess noise in a single detector. In principle, to conclude that a SGWB is really present one should exclude that the excess noise is not due to some source of noise not taken into account. The difficulty in doing this is also increased by the fact that the  signal-to-noise ratio does not increase with the observation time, differently from what happens in the search for continuous signals. On the other hand, the signal would show up as a coherent stochastic process between two or more detectors.  
Then, a typical analysis strategy consists in cross-correlating the data from multiple detectors. By indicating with $s_1(t),~s_2(t)$ the data streams from two detectors, the cross-correlation is
\be
Y=\int_{-T_{obs}/2}^{+T_{obs}/2}dt \int_{-T_{obs}/2}^{+T_{obs}/2}dt'~s_1(t)s_2(t')Q(t-t')
\ee
where $Q(t-t')$ is a filter function chosen to maximize the signal-to-noise ratio. In the frequency domain the optimal filter function takes the form
\be
\tilde{Q}(f)\propto \frac{\Gamma(f) \Omega_{gw}(f)}{f^3S_1(f)S_2(f)}
\ee
where $S_1,S_2$ are the power spectral noise density of the two detectors and $\Gamma(f)$ is the overlap reduction function which takes into account the fact that the two detectors can see a different signal because they are at a different location or because they have a different angular sensitivity. In particular, if the separation between the two detectors is much larger than the signal reduced wavelength, the correlation is strongly suppressed. On the other hand, if the distance among two detectors is very small, or if they are co-located, the identification of coherent disturbances is not a trivial task. The sensitivity of a pair of detectors is usually given in terms of the minimum detectable amplitude for a flat spectrum~\cite{alle}:
\be
\Omega_{min}\approx \frac{34}{H_0^2\sqrt{T_{obs}}}\left [\int_0^{\infty}\frac{\Gamma^2(f)}{f^6S_1(f)S_2(f)}df\right]^{-1/2}
\ee 
where a false alarm rate of $1\%$ and a false dismissal rate of $10\%$ have been considered.

\begin{figure}
\begin{center}
\includegraphics[width=0.6\textwidth]{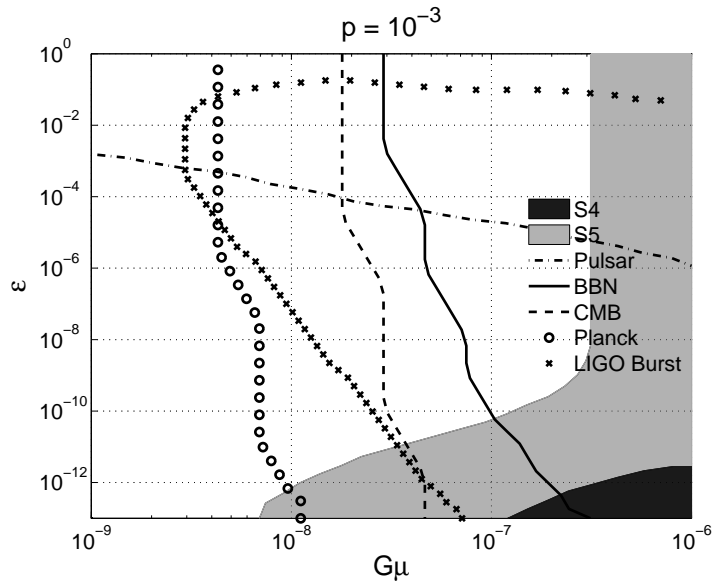}
\caption{Exclusion regions in the $\epsilon-G\mu$ plane. LIGO-S5 results exclude region of low $\epsilon$ and low $G\mu$ (Reprinted from Nature~\protect \cite{natS5}).   
\label{fig:stoch_cs}}
\end{center}
\end{figure}

The full-S5 data set from LIGO detectors has been analyzed to search for a SGWB cross-correlating data from the detector pairs H1-L1 and H2-L1~\cite{natS5}. The effective observation time was $\sim 293$ days. The analysis was focused on the frequency band $41-170$ Hz, which includes about $99\%$ of the instrumental sensitivity. A bayesian $95\%$ degree of belief upper limit has been set, taking the S4 posterior as a prior:  
\be
\Omega_{gw}(f)<6.9\cdot 10^{-6}
\label{ome}
\ee
Models to explain the element abundance observation constrain total energy at the time of the Big-Bang nucleosynthesis to $\int \Omega_{gw}(f) d(ln~f)<1.5\cdot 10^{-5}$. Allocating all the energy to the analyzed frequency band implies an upper bound of $1.1\cdot 10^{-5}$. 
Then the limit of Eq. \ref{ome} beats the indirect limits provided by Big-bang nucleosynthesis and cosmic microwave background, see Fig.\ref{fig:stoch_new}. This result constrains also models of cosmic super-strings, in which the gravitational background is due to the superposition of many cusp burst signals~\cite{xavi}, excluding regions in the $G\mu-\epsilon$ plane, as shown in Fig.\ref{fig:stoch_cs}.  

The analysis of data from the pair H1-H2 is also underway. From one hand having two co-located detectors would give a sensitivity improvement of about one order of magnitude. On the other, the presence of correlations between the two data streams reduces the gain. A big effort is being done in order to identify all the environmental contributions to the H1-H2 cross-correlation. Also the search for non-isotropic backgrounds is being considered, and an analysis method optimized for this kind of signal, called {\it radiometer} analysis~\cite{rad}, has already produced results for S4 and is being applied to S5 data.   

Advanced detectors should push the upper limit a couple of orders of magnitude below the current limit thus further constraining the parameter space of various models of cosmological SGWB, see Fig.\ref{fig:stoch_new}. In particular, for cosmic super-string models they could exclude regions of $G\mu>10^{-11}$ and $\epsilon>10^{-10}$. They should be also able to put constraints on Pre-Big-Bang models by excluding regions in the $f_1-\mu$ plane. 
 
\section{Conclusions}
The search for continuous gravitational wave signals and stochastic gravitational wave backgrounds in LIGO and Virgo data has already produced several upper limits of astrophysical interest, altough no detection. New results will come soon by analyzing most recent data and using improved analysis pipelines. The development of more sensitive and robust methods will follow in the next years in order to be ready for the advanced detectors era.

\section*{Acknowledgments}
The authors gratefully acknowledge the support of the United States
National Science Foundation for the construction and operation of the
LIGO Laboratory, the Science and Technology Facilities Council of the
United Kingdom, the Max-Planck-Society, and the State of
Niedersachsen/Germany for support of the construction and operation of
the GEO600 detector, and the Italian Istituto Nazionale di Fisica
Nucleare and the French Centre National de la Recherche Scientifique
for the construction and operation of the Virgo detector. The authors
also gratefully acknowledge the support of the research by these
agencies and by the Australian Research Council, 
the International Science Linkages program of the Commonwealth of Australia,
the Council of Scientific and Industrial Research of India, 
the Istituto Nazionale di Fisica Nucleare of Italy, 
the Spanish Ministerio de Educaci\'on y Ciencia, 
the Conselleria d'Economia Hisenda i Innovaci\'o of the
Govern de les Illes Balears, the Foundation for Fundamental Research
on Matter supported by the Netherlands Organisation for Scientific Research, 
the Polish Ministry of Science and Higher Education, the FOCUS
Programme of Foundation for Polish Science,
the Royal Society, the Scottish Funding Council, the
Scottish Universities Physics Alliance, The National Aeronautics and
Space Administration, the Carnegie Trust, the Leverhulme Trust, the
David and Lucile Packard Foundation, the Research Corporation, and
the Alfred P. Sloan Foundation.

\end{document}